\documentclass[11pt]{article}
\usepackage[sfdefault]{carlito} 
\usepackage{geometry}
\usepackage{setspace}
\usepackage{titlesec}
\PassOptionsToPackage{hyphens}{url}\usepackage{hyperref}
\usepackage{graphicx}
\usepackage{amsmath}
\usepackage{bm}
\usepackage{wrapfig}
\usepackage{xcolor}
\usepackage{authblk}
\usepackage[export]{adjustbox}
\usepackage{subcaption}
\usepackage{tikz} 
\usetikzlibrary{matrix}

\usepackage{comment}

\titleformat{\section}{\normalfont\fontsize{11}{13}\bfseries\sffamily}{\thesection}{1em}{}
\titleformat{\subsection}{\normalfont\fontsize{11}{13}\bfseries\sffamily}{\thesubsection}{1em}{}
\titleformat{\title}{\normalfont\fontsize{14}{16}\bfseries\sffamily}{}{0em}{}

\title{Coupled-cluster theory for the ground state and for excitations}

\author[2]{Andreas Grüneis}
\author[1]{\textbf{Evgeny Moerman}\footnote{Coordinator of the contribution} }
\author[1]{\textbf{Matthias Scheffler}\textsuperscript{* }}
\author[3]{Tonghao Shen}
\author[3]{Igor Ying Zhang}

\affil[1]{NOMAD Laboratory at the FHI of the Max Planck Society, Faradayweg 4-6, 14195 Berlin, Germany}
\affil[2]{Institute for Theoretical Physics, TU Wien, Wiedner Hauptstra{\ss}e 8--10/136, 1040 Vienna, Austria}
\affil[3]{Department of Chemistry, Fudan University
, Shanghai 200433, People’s Republic of China}

\date{}

\newlength{\tempdima}
\newcommand{\rowname}[1]
{\rotatebox{90}{\makebox[\tempdima][c]{\textbf{#1}}}}

\begin{document}

\maketitle

\section*{Summary}

In the molecular quantum chemistry community, coupled-cluster (CC) methods are well-recognized for
their systematic convergence and reliability. The extension of 
the theory to extended systems has been comparably recent~\cite{zhang2019coupled}, so that developments and studies of periodic CC methods for both the ground-state and for excited states are still active fields 
of research and provide valuable benchmark data when the
reliability of density functional approximations is questionable. In this contribution we describe the CC-aims interface between
the FHI-aims and the Cc4s software packages. 
This linkage makes a variety of correlated wave function-based ground-state methods including 
M\o ller-Plesset
perturbation theory (MP2), the random-phase
approximation (RPA) and the
\emph{gold-standard of quantum chemistry} CCSD(T)
method for both molecular and periodic applications accessible.
This contribution discusses these ground-state methods for clusters and molecules, as 
well as for periodic systems. In particular, we discuss recent advancements and the
implementation of the equation-of-motion CC method
for the calculation of ionization (IP-EOM-CCSD) and
electron attachment (EA-EOM-CCSD) processes.
Open questions and routes to solutions are discussed as well.

\section*{Current Status of the Implementation}

\begin{figure}
    \centering
    \includegraphics[width=\linewidth]{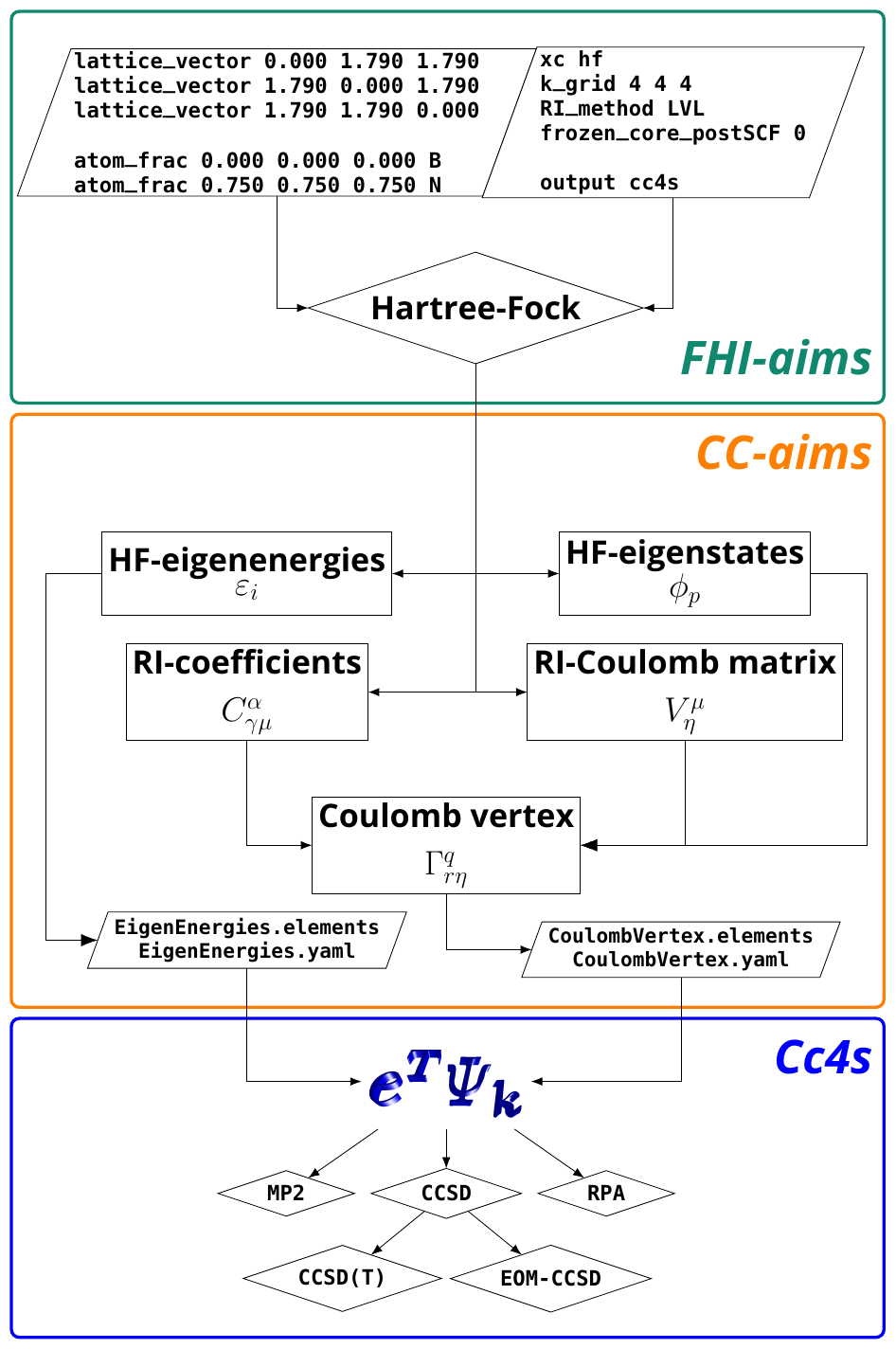}
    \caption{The Cc4s@FHI-aims workflow}
    \label{fig:workflow}
\end{figure}

The Cc4s code constitutes an open-source quantum chemistry software package, which features several correlated wave function methods. As a post-SCF code, Cc4s requires single-particle
eigenergies and wave functions from a mean-field
calculation, which another electronic structure
package must provide. Using the CC-aims interface~\cite{moerman2022interface}
the relevant quantities are conveniently obtained from 
a FHI-aims Hartree-Fock calculation. CC-aims, then, parses and converts the
needed quantities
(e.g the single-particle eigenergies) 
to a Cc4s-compatible format and
computes additional quantities 
(e.g the Coulomb vertex). The files generated by CC-aims
are then used to launch a Cc4s calculation with any of the therein implemented
wave-function methods. This workflow is illustrated in Figure \ref{fig:workflow}.

Below, we discuss the coupled-cluster approach for the electronic ground-state
and for excited states for molecules and clusters and for periodic solids.

In practice, the limiting factor of CC methods
is the substantial memory requirement, which results from the size of the Coulomb integral tensor

\begin{equation}
     V^{pq}_{rs} = \int d\bm{r}\bm{r}'\, 
    \frac{\phi_p^{*}({\bm{r}})\phi_q^{*}(\bm{r'})\phi_r({\bm{r}})\phi_s(\bm{r'})}{|\bm{r} - \bm{r'}|}\text{,}
\end{equation}

where $\phi_p$, $\phi_q$, $\phi_r$ and $\phi_s$
denote four single-particle wave functions.
As a consequence, the memory to store
this tensor scales with the fourth power of 
the basis set size. By performing a low-rank
decomposition of $V^{pq}_{rs}$, the Coulomb vertex
$\Gamma_{r}^{p,\eta}$,
a rank three tensor with a significantly smaller memory footprint~\cite{hummel2017low} is defined

\begin{equation}
    V^{pq}_{rs} \approx
    \sum_{\eta}
    \Gamma_{r}^{p\eta} \Gamma^{q*}_{s\eta}\text{,}
\end{equation}

where $\eta$ indexes the basis functions of an
auxiliary basis. As a memory-saving measure,
CC-aims computes the Coulomb vertex. The auxiliary basis is that of the resolution-of-the-identity (RI) employed in FHI-aims. In combination with CC-aims, one can choose between the RI-V scheme for molecular
applications and its localized approximation, the RI-LVL scheme for periodic
ones. While the former is deemed very
accurate, the latter is more memory-efficient
but in general of insufficient accuracy~\cite{ihrig2015accurate}. 
However, the incompleteness of the RI-LVL auxiliary basis for Hartree-Fock, MP2 and RPA methods can be satisfactorily resolved by manually adding a few auxiliary $f$ -, $g$- and $h$-type basis functions with small eﬀective charges~\cite{ihrig2015accurate}. This approach is also applicable for CC calculations.
The memory footprint of the Coulomb vertex can be further reduced by performing 
a principal component analysis, with which
up to 70\% of the Coulomb vertex can be discarded
in many applications~\cite{hummel2017low}.

Currently, the periodic CC infrastructure of FHI-aims does not allow to perform spin-polarized calculations and the starting point needs to be canonical Hartree-Fock orbitals. For periodic applications, the most important constraint is given by the inability of 
Cc4s to perform an explicit $k$-point summation. 
Instead, Cc4s requires a super cell based
treatment of a periodic system. 
As a consequence, Cc4s does not make use of the
translational symmetry of crystalline systems,
which would reduce 
the memory scaling by a factor of $N_k$
and the computational scaling by a factor of $N_k^2$, with $N_k$ being the number of $k$-points.

\section*{Usability and Tutorials}

\begin{figure}[htbp]
\centering
 \begin{tikzpicture}   
 \matrix (fig) [matrix of nodes]{
 \includegraphics[width=0.47\linewidth]{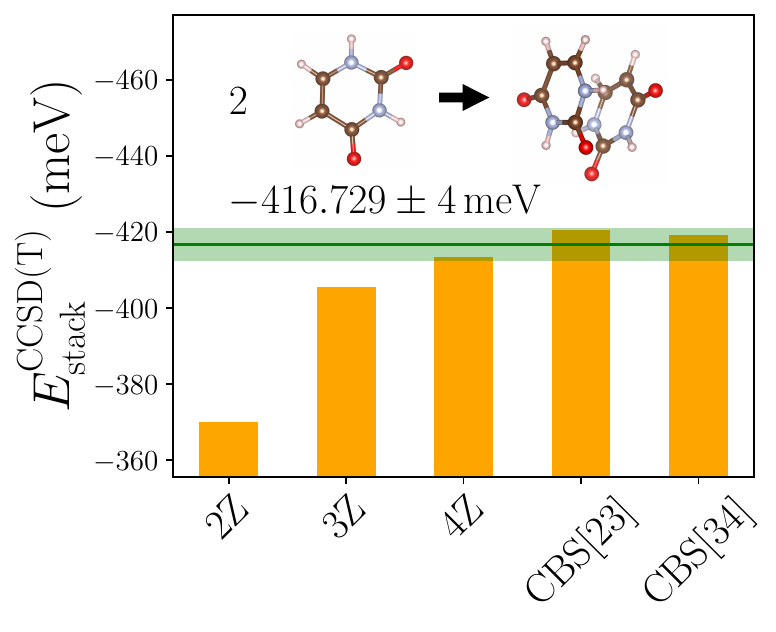}&
 \vspace{1cm}
 \raisebox{0.46cm}{\includegraphics[width=0.47\linewidth]
 {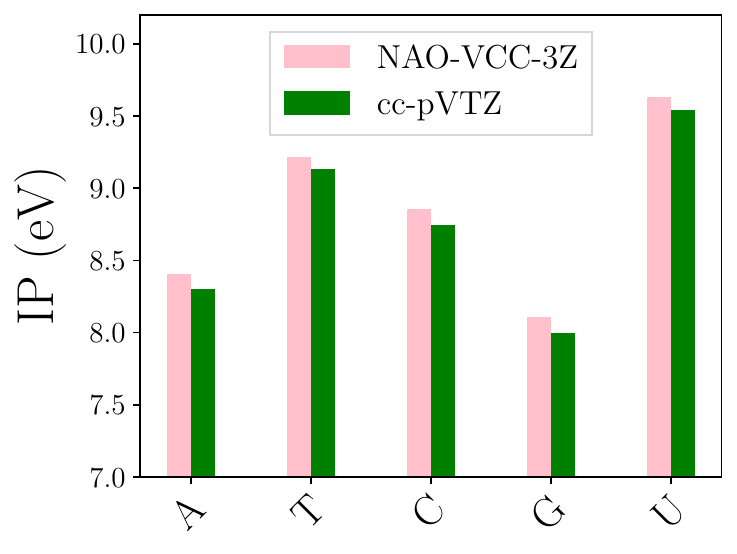}}
 \\[0cm] 
 |[text width=1.85in]| {\subcaption{}\label{fig:1a}}
 &
 |[text width=1.85in]| {\subcaption{}\label{fig:1b}}
 \\[0.0cm] 
 \includegraphics[width=0.47\linewidth]
 {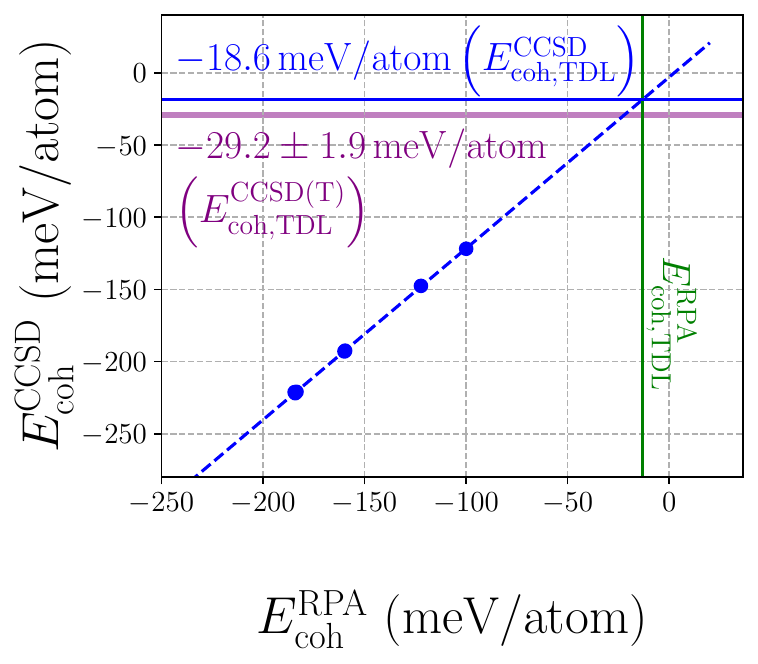}
 &
 \includegraphics[width=0.47\linewidth]{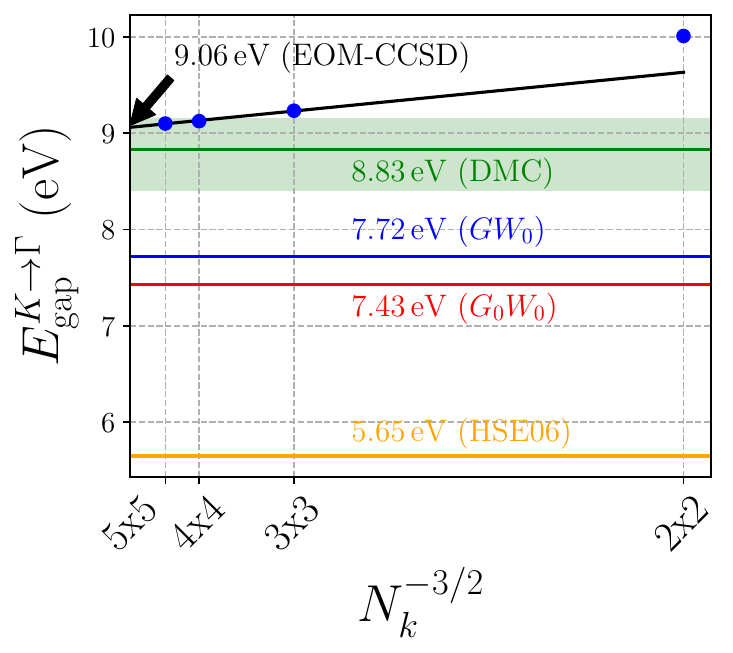}
 \\[0cm] 
 |[text width=1.85in]| {\subcaption{}\label{fig:1c}}
 &
 |[text width=1.85in]| {\subcaption{}\label{fig:1d}}
 \\
  };
  
  \draw[-latex, white] ([xshift=-2mm,yshift=2mm]fig-1-1.south west)  -- ([xshift=-2mm,yshift=-2mm]fig-1-1.north west) node[midway,above,sloped, black]{Molecules};

 \draw[-latex, white] ([xshift=-2mm,yshift=2mm]fig-3-1.south west)  -- ([xshift=-2mm,yshift=-2mm]fig-3-1.north west) node[midway,above,sloped, black]{Extended solids};
 \draw[-latex, white] ([yshift=2mm,xshift=2mm]fig-1-1.north west)  -- ([yshift=2mm,xshift=-2mm]fig-1-1.north east) node[midway,above, black]{Ground state};
 \draw[-latex, white] ([yshift=7.5mm,xshift=2mm]fig-1-2.north west)  -- ([yshift=7.5mm,xshift=-2mm]fig-1-2.north east) node[midway,below, black]{Excited state};
 \end{tikzpicture}
    \caption{A selection of molecular and periodic
CC calculations with the NAO-VCC-$n$Z basis sets:
Figure \ref{fig:1a} 
stacking energy of a uracil dimer on the CCSD(T) level of theory compared to Reference~\cite{al2021interactions},
Figure \ref{fig:1b} 
vertical ionization potentials of adenine (A), thymine (T), cytosine (C), guanine (G)
and uracil (U) via IP-EOM-CCSD compared to Reference~\cite{tripathi2022performance}, 
Figure \ref{fig:1c} 
Finite-size convergence of the cohesive energy
of Neon using the CCSD method plotted against the RPA results.
The (T)-correction has been found to be mostly system-size-independent
and has been added to the extrapolated CCSD result.
Figure \ref{fig:1d} 
Electronic band gap of two-dimensional boron nitride via IP- and EA-EOM-CCSD compared to HSE06 and higher-level correlated methods in Reference~\cite{hunt2020diffusion}. The opaque green area illustrates
the uncertainty of the DMC result.}
    \label{fig:cc-use-cases}
\end{figure}

A variety of FHI-aims based calculations with the CCSD, CCSD(T) and the EOM-CCSD implementations in Cc4s have already 
been performed for small and medium-sized molecules
and for crystalline insulators and semiconductors.
Tutorials for using the workflow involving
FHI-aims, CC-aims and Cc4s for both molecules
and crystalline materials are available on
the FHI-aims platform~\cite{aimscctutorial}.
These tutorials include a detailed step-by-step guide on the installation of the
necessary software (i.e CC-aims and Cc4s) and the calculation of
ground-state and excitation energies. Currently, the tutorial consists of 
three parts. The first one introduces the FHI-aims/CC-aims/Cc4s workflow
by an exemplary calculation of the MP2 correlation energy of the paracetamol
molecule. The second tutorial focuses on a periodic application and demonstrates
the importance of a careful consideration of the system-size dependent error
for correlated wave function methods. To that end, the necessary steps involving
the calculation of the CCSD cohesive energy of Neon are presented, the results
of which are also shown in Figure \ref{fig:1c}. In the third tutorial, 
the additional steps to compute quasi-particle energies via the 
IP- and EA-EOM-CCSD method are described in detail, for which the
ionization potential of uracil (see Figure \ref{fig:1b})
and the band gap of \textit{trans}-polyacetylene is computed~\cite{moerman2024finite}.

Figure \ref{fig:cc-use-cases} shows four typical 
applications of CC theory using FHI-aims for molecular
and extended system involving both the ground state
and charged excited states. Figure \ref{fig:1a}
and \ref{fig:1b} confirm the accuracy of both ground state 
and excited state properties for molecules using our 
valence-correlation
consistent NAO-VCC-$n$Z basis sets~\cite{zhang2013numeric}. Via extrapolation to the complete basis set (CBS) limit, we are able to reproduce the stacking energy of uracil (Figure \ref{fig:1a}) within a few
meV. 
Going beyond the ground-state, in Figure \ref{fig:1b} we show results 
of the IP-EOM-CCSD method to obtain the ionization potentials
of the five nucleobases, which are in good agreement (within $\approx 100\,\text{meV}$) with previously reported
values by Tripathi et al.~\cite{tripathi2022performance}. 

The remaining deviation stems from the use of different finite basis sets: cc-pVTZ in Reference~\cite{tripathi2022performance} and NAO-VCC-3Z in our study. For ground-state properties, we have found that the NAO-VCC-$n$Z
basis set performs better than the cc-pV$n$Z  basis set for advanced correlation methods, including MP2, RPA, and CCSD(T)~\cite{zhang2013numeric, shen2019massive}. This discrepancy can be further minimized by extrapolating to the CBS limit.

The remaining deviation is attributable to the utilization of a 3Z basis set, for which 
a remaining difference of 
that magnitude to the CBS limit is expected.

Figure \ref{fig:1d} demonstrates the 
applicability of the  EOM-CCSD method to obtain
quasi-particle band gaps. For that we studied the
fundamental, indirect $K\to\Gamma$ band gap of two-dimensional
hexagonal boron nitride (hBN), for which different state-of-the-art
methods like generalized KS-DFT and the $GW$ approximation and highly accurate methods like
Diffusion Monte Carlo (DMC) yielded very different results. Our analysis
shows good agreement with the DMC result by 
Hunt et al.~\cite{hunt2020diffusion}, confirming that an accurate
treatment of electronic correlation is crucial in this system.
Figure \ref{fig:1c} shows the cohesive energy of the Neon crystal.
The stability of this crystal is almost exclusively determined by 
van-der-Waals interactions, which CC methods are known to precisely
capture. 
With Cc4s@FHI-aims we obtain a cohesive energy of 
$-17.9\,\text{meV/atom}$ at the CCSD level. 
The experimental value, already adjusted for zero-point fluctuations, is $-27\,\text{meV/atom}$, with an estimated zero-point contribution of $7\,\text{meV/atom}$~\cite{rosciszewski1999ab}.
For these calculations the NAO-VCC-2Z and -3Z basis sets were employed. Performing a conventional two-point extrapolation 
to the CBS limit yields the results in Figure \ref{fig:1c} 
with a remaining difference to the CBS limit of $8\,\text{meV/atom}$, which was added to all
of the data points. A more accurate cohesive energy was obtained by accounting for triple excitations in a 
perturbative manner. That triple (T)-correction was found
to be mostly independent of the system size and was
determined to lie between $-8.7\,\text{meV/atom}$
and $-12.6\,\text{meV/atom}$. By adding these values
to the extrapolated CCSD cohesive energy, a CCSD(T) 
result of $-29.2\pm1.9\,\text{meV/atom}$ was found,
which is in excellent agreement with the experimental 
finding.

\section*{Future Plans and Challenges}

As mentioned before, the currently biggest limitation in performing
periodic CC calculations via Cc4s is the lack of an explicit $k$-point
summation. 
The translational symmetry of a periodic system is reflected in the
block-sparsity of the CC tensors. A block-sparse implementation of the
tensor contraction engine in Cc4s is currently under development and is
expected to severely reduce the computational and memory requirements of 
CC calculations.
Further future plans involve the extension of the Cc4s functionalities to include spin-polarized and non-HF based calculations. 
The convergence of CC ground and excited state properties with respect to system size can be very slow. As has been in part demonstrated in Figure \ref{fig:1c}, one can correct the corresponding finite-size errors by additionally employing another, computationally cheaper non-CC method (e.g the RPA for the ground-state or the $G_0W_0$ method for quasi-particles), which exhibits a similar system size convergence. However, via the implementation of a block-sparse tensor treatment in Cc4s, more work has to be done to verify the accuracy and the limitations of this methodology.
Development of new strategies to reduce the basis set incompleteness error is the second, very important future goal,
to improve the precision of CC calculations. Particularly, in ground-state CC applications, where the quantity of interest itself is sometimes
on the order of a few meV, one is forced to use very big basis
sets even if extrapolation schemes are accessible to achieve the necessary precision. 
One solution is to use the much more compact natural orbital basis instead of the 
canonical HF basis~\cite{lowdin1955quantum}. One way to construct natural orbitals for correlated wave function methods is to construct the one-electron reduced density matrix at the MP2 level~\cite{grüneis2011natural}. Diagonalization of the density matrix yields 
the natural orbitals and selection of only those with a sufficiently high
occupation numbers allows to significantly reduce the number of 
single-particle states involved in the subsequent CC calculation,
thus reducing the overall computational cost. The possibility to construct natural orbitals in FHI-aims or CC-aims is therefore a very desirable
feature.

\section*{Acknowledgements}
We acknowledge the support by TEC1p [the European Research Council (ERC) Horizon 2020 research 
  and innovation program, Grant Agreement No.740233].



\begin{thebibliography}{10}

\bibitem{zhang2019coupled}
Igor Ying Zhang, Andreas Grüneis, \textit{Coupled Cluster Theory in Materials Science}. Front. Mater. \textbf{6}, 123 (2019)

\bibitem{moerman2022interface}
Evgeny Moerman, Felix Hummel, Andreas Grüneis, Andreas Irmler,
and Matthias Scheffler, \textit{Interface to high-performance periodic coupled-cluster theory calculations with atom-centered, localized basis functions}. J. Open Source Softw.
\textbf{7}, 4040 (2022)

\bibitem{hummel2017low}
Felix Hummel, Theodoros Tsatsoulis, , Andreas Grüneis, 
\textit{Low rank factorization of the Coulomb integrals for periodic coupled cluster theory}. J. Chem. Phys. \textbf{146}, 124105 (2017)

\bibitem{ihrig2015accurate}
Arvid Conrad Ihrig, Jürgen Wieferink, Igor Ying Zhang, Matti Ropo, Xinguo Ren, Patrick Rinke,
Matthias Scheffler and Volker Blum, \textit{Accurate localized resolution of identity approach for linear-scaling
hybrid density functionals and for many-body perturbation theory}.
New J. Phys. \textbf{17}, 093020 (2015)


\bibitem{aimscctutorial}
The coupled-cluster tutorial of FHI-aims. \url{https://periodic-cc-methods-in-fhi-aims-fhi-aims-club-tu-9d435dcb75ed8d.gitlab.io/}

\bibitem{moerman2024finite}
Evgeny Moerman, Alejandro Gallo, Andreas Irmler, Tobias Schäfer, Felix Hummel, Andreas Grüneis, Matthias Scheffler, \textit{Finite-size Effects in periodic EOM-CCSD for Ionization Energies and Electron Affinities: Convergence Rate and Extrapolation to the Thermodynamic Limit}. arXiv:2409.03721 (2024)

\bibitem{zhang2013numeric}
Igor Ying Zhang, Xinguo Ren, Patrick Rinke, Volker Blum, Matthias Scheffler, \textit{Numeric atom-centered-orbital basis sets with valence-correlation consistency from H to Ar}. New J. Phys. 15 123033 (2013)

\bibitem{lowdin1955quantum}
Per-Olov Löwdin, \textit{Quantum Theory of Many-Particle Systems. I. Physical Interpretations by Means of
Density Matrices, Natural Spin-Orbitals, and Convergence Problems
in the Method of Configurational Interaction}. Phys. Rev. \textbf{97}, 
1474 (1955)

\bibitem{grüneis2011natural}
Andreas Grüneis, George H. Booth, Martijn Marsman, James Spencer, Ali Alavi, Georg Kresse, \textit{Natural Orbitals for Wave Function Based Correlated Calculations
Using a Plane Wave Basis Set}.  J. Chem. Theory Comput. \textbf{7}, 2780–2785 (2011)

\bibitem{al2021interactions}
 Yasmine S. Al-Hamdani, P{\'e}ter R. Nagy,  Andrea Zen, Dennis Barton, Mih{\'a}ly K{\'a}llay, Jan Gerit Brandenburg,  Alexandre Tkatchenko, \textit{Interactions between large molecules pose a puzzle for reference quantum mechanical methods}. Nat. Commun. \textbf{12}, 3927 (2021)

\bibitem{tripathi2022performance}
Divya Tripathi, Achintya Kumar Dutta, \textit{The performance of approximate EOM-CCSD for ionization potential and electron affinity of genetic material subunits: A benchmark investigation}. Int. J. Quantum Chem. \textbf{122}, e26918 (2022)

\bibitem{shen2019massive}
Tonghao Shen, Zhenyu Zhu, Igor Ying Zhang, Matthias Scheﬄer,
\textit{Massive-parallel implementation of the resolution-of-identity coupled-cluster approaches in the numeric atom-centered orbital framework for molecular systems}.
J. Chem. Theory Comput. , \textbf{15}, 4721–4734 (2019)


\bibitem{hunt2020diffusion}
 R. J. Hunt, B. Monserrat,  V. Zólyomi, and N. D. Drummond, \textit{Diffusion quantum Monte Carlo and GW study of the electronic properties of monolayer and bulk
hexagonal boron nitride}. Phys. Rev. B \textbf{101}, 205115 (2020)

\bibitem{rosciszewski1999ab}
Krzysztof Rościszewski, Beate Paulus, Peter Fulde, and Hermann Stoll,
\textit{Ab initio calculation of ground-state properties of rare-gas crystals}. Phys. Rev. B \textbf{60}, 7905 (1999)

\bibitem{gruber2018applying}
Thomas Gruber, Ke Liao, and Theodoros Tsatsoulis, Felix Hummel, Andreas Grüneis, \textit{Applying the Coupled-Cluster Ansatz to Solids and Surfaces in the Thermodynamic Limit}. Phys. Rev. X \textbf{8}, 021043 (2018)

\bibitem{moerman2024tobepublished}
Evgeny Moerman,  Alejandro Gallo, Andreas Irmler, 
Felix Hummel, Andreas Grüneis,
and Matthias Scheffler, to be published


  
\end{thebibliography}
\end{document}